\documentclass[twocolumn,aps,prl,showpacs,preprintnumbers]{revtex4}%
\usepackage{mathrsfs}
\usepackage{amsmath}
\usepackage{amsfonts}
\usepackage{amssymb}
\usepackage{amsthm}
\usepackage{graphicx}
\usepackage{color}
\usepackage{txfonts}%
\providecommand{\U}[1]{\protect\rule{.1in}{.1in}}
\begin{document}
\title{Thermodynamic Magnon Recoil for Domain Wall Motion}
\author{Peng Yan$^{1}$}
\author{Yunshan Cao$^{2}$}
\author{Jairo Sinova$^{1}$}
\affiliation{$^{1}$Institut f\"{u}r Physik, Johannes Gutenberg Universit\"{a}t Mainz,
Staudinger Weg 7, 55128 Mainz, Germany}
\affiliation{$^{2}$Kavli Institute of NanoScience, Delft University of Technology,
Lorentzweg 1, 2628 CJ Delft, The Netherlands}

\begin{abstract}
We predict a thermodynamic magnon recoil effect for domain wall motions in the presence of temperature
gradients. %The theory incorporates non-equilibrium thermodynamics .
All current thermodynamic theories assert that a magnetic domain
wall must move toward the hotter side, %in the presence of a temperature gradient.
%The arguments are
based on equilibrium thermodynamic arguments.
%which is not
%strictly the case in the presence of a temperature gradient. Moreover,
Microscopic calculations on the other hand show that a domain wall can move either along or
against the direction of heat currents, depending on how strong the heat currents
are reflected by the domain wall. We have resolved the inconsistency between these two approaches by augmenting
the theory in the presence of thermal gradients by incorporating in the %to incorporate non-equilibrium thermodynamic components,
free energy of domain walls by a heat current term present in
non-equilibrium steady states. The condition to observe a domain wall
propagation toward the colder regime is derived analytically and can be tested
by future experiments.

\end{abstract}

\maketitle

%75.60.Jk Magnetization reversal mechanisms\\
%75.75.+a Magnetic properties of nanostructures \\
%75.60.Ch Domain walls and domain structure
%85.75.-d Magnetoelectronics; spintronics:
%devices exploiting spin polarized transport or integrated magnetic fields
%85.70.Ay Magnetic device characterization, design and modeling
%75.30.Ds Spin waves\\
%----------------------------------------------------------------%
Spincaloritronics is the subfield of spintronics which explores spin-dependent
phenomena coupled to thermal gradients \cite{Bauer}. A very important question
within this field is how a magnetic domain wall (DW) can move under a
temperature gradient. This question has attracted much attention owing to its
applicability in magnetic insulators \cite{Nowak1,Nowak2,XR,Kovalev,Jiang} for
potential applications in logic devices \cite{Cowburn} and data storage
technology \cite{Parkin}. The conventional approach using static magnetic
fields \cite{Walker} is well established with high DW velocities
\cite{Atkinson,Beach,Hayashi}, but does not allow for the synchronous motion
of multiple domain walls. Synchronous current-induced domain wall motion due
to spin-transfer torques \cite{BergerSlon,Zhang,Gen} and/or spin-orbit torques
\cite{Miron,Liu,Fert,Koopmans,Emori} formed an alternative way to efficiently
manipulate the magnetization configuration, but the required high-current
densities cause problems, such as Joule heating owing to Ohmic losses. Heat
itself has been proposed as an efficient control parameter to overcome the
problems during the emergence of spincaloritronics \cite{Hatami}.
Fully understanding and predicting new controlled ways to move the domain walls by
magnonic heat currents is paramount to exploiting fully all their future device possibilities.

There is at present an theoretical incomplete understanding of  temperature-gradient-driven DW motion.
There are two types of theories, i.e., a macroscopically
thermodynamic theory \cite{Nowak1,Nowak2,XR,Kovalev,Jiang} and a
microscopically magnonic one \cite{Yan1,Yan2,Wang}. The theories contradict
each other in certain regimes.
In previous thermodynamic theories \cite{Nowak1,Nowak2,XR}, a magnetic domain
wall at finite temperature $T$ is treated as a thermodynamic object with
free energy,
%\begin{equation}
$F=U-TS$, %\label{Free}%
%\end{equation}
where $U$ is its internal energy and $S$ is its entropy. The free energy of
the DW can also be expressed as the difference between a system with a DW
minus that of the same system without the wall
%\begin{equation}
$\Delta F=\Delta U-T\Delta S$. %\label{FreeDiff}%
%\end{equation}
Thermodynamic calculations \cite{Nowak1,Nowak2,XR} show that, far below the
Curie temperature, the entropy $\Delta S\left(  T\right)  $ increases and the
free energy $\Delta F\left(  T\right)  $ decreases with the temperature. This
leads to a conclusion that the DW must move towards regions with higher
temperatures due to the entropic force, with the propagation velocity
proportional to the temperature gradient $\nabla T$
\cite{Nowak1,Nowak2,XR,Kovalev}.
%which seems to be consistent with the only
%experiment so far
This tendency has been observed in experiments \cite{Jiang}.

On the other hand, there are microscopic angular momentum transfer \cite{Yan1}
and linear momentum transfer \cite{Yan2} theories for the magnon-driven DW
motion. The DW moves along the opposite direction of current flows if the
former mechanism dominates \cite{Kovalev,Yan1}, while along the same direction
when the latter one is more important \cite{Yan2,Wang,Kim}, i.e., there are
strong spin-wave reflections by the wall \cite{Yan2,Yan3}. The proposed
mechanism of magnonic linear momentum transfer has been confirmed in various
systems including ferromagnets \cite{Yan2,Wang}, antiferromagnets
\cite{Erlend,Se}, and spin textures with Dzyaloshiskii-Moriya interaction
\cite{Weiwei}.

In this Letter, we show that the inconsistency between these two types of
theories can be resolved by augmenting the free energy by a term from the heat
current which always flows in a non-equilibrium steady state in the presence
of a temperature gradient. The heat current gets modulations by the DW with
momentum-conserving back-scatterings. It then leads to a force that pushes a DW
%propagation
 toward the colder region. We predict a new thermodynamic
magnon recoil effect for the domain wall motion in temperature gradients.
Under conditions of a strong backscattering, a high magnon thermal
conductivity, and a slow magnon group velocity, this magnon recoil
effect surpasses the previously identified entropic force
\cite{Nowak1,Nowak2,XR,Kovalev}. Such a regime can be achieved in yttrium iron
garnet (YIG) and other
ferromagnetic insulators, as we show below.
\begin{figure}[ptbh]
\begin{center}
\includegraphics[width=8cm]{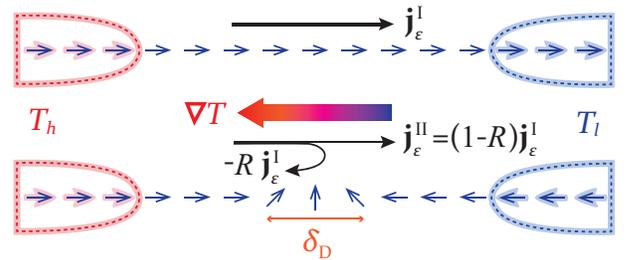}
\end{center}
\caption{(Color online) Heat currents in a 1D magnetic wire in the presence of
a temperature gradient $\nabla T$ without (upper panel) and with (lower panel)
a domain wall with width $\delta_{D}$, denoted by $\mathbf{j}_{\varepsilon
}^{\text{I}}$ and $\mathbf{j}_{\varepsilon}^{\text{II}},$ respectively. The
presence of the domain wall causes the reflection of heat currents with a
probability $R$.}%
\end{figure}

We consider a one dimensional (1D) magnetic wire connecting two thermal
reservoirs with temperatures $T_{h}$ and $T_{l}$ $\left(  T_{h}>T_{l}\right)
$ as shown in Fig. 1. The temperature gradient $\nabla T$ then drives magnon
heat currents $\mathbf{j}_{\varepsilon}^{\text{I}}$ and $\mathbf{j}%
_{\varepsilon}^{\text{II}}$ in a uniformly magnetized wire (as shown in the
upper panel in Fig. 1) and in a wire with a domain wall (shown in the lower panel
in Fig. 1), respectively, at non-equilibrium steady states. The presence of
the domain wall causes finite magnon reflections \cite{Yan3}, thereby reducing
the heat current, i.e., $\mathbf{j}_{\varepsilon}^{\text{II}}\leqslant
\mathbf{j}_{\varepsilon}^{\text{I}}.$ They are equal only when the potential
generated by the domain wall is reflectionless \cite{Yan1,Yan2}. A heat
current $\mathbf{j}_{\varepsilon}^{\text{I,II}}$ can modify the rate of the
change of the entropy \cite{Landau,Tatara}%
\begin{equation}
\frac{dS^{\text{I,II}}}{dt}=-\int dx\mathbf{j}_{\varepsilon}^{\text{I,II}%
}\cdot\frac{\nabla T}{T^{2}}, \label{Rate}%
\end{equation}
and this entropy change modifies the free energy. We therefore write the free
energy of the wall as%
\begin{equation}
\Delta F=\Delta F_{\text{e}}+\Delta F_{\text{ne}}. \label{Totalfree}%
\end{equation}
Here $\Delta F_{\text{e}}$ is the equilibrium part as treated in previous
thermodynamic theories \cite{Nowak1,Nowak2,XR}, which may originate from the
modified magnon density of states due to the wall \cite{Yan3,Vindigni,XR},
while $\Delta F_{\text{ne}}$ is the non-equilibrium part proportional to the
heat current. Its change rate is then%
\begin{align}
\frac{d\left(  \Delta F_{\text{ne}}\right)  }{dt}  &  =-\int T\left(  \dot
{S}^{\text{II}}-\dot{S}^{\text{I}}\right)  dx\nonumber\\
&  =\int R\left(  -\mathbf{j}_{\varepsilon}^{\text{I}}\cdot\frac{\nabla T}%
{T}\right)  dx, \label{NEfree}%
\end{align}
where we introduced a parameter $R\left(  T\right)  $ denoting the
temperature-dependent reflection probability of magnon heat currents by the
wall. The magnon heat current in the presence of the domain wall is thus
modified, i.e., $\mathbf{j}_{\varepsilon}^{\text{II}}=\left(  1-R\right)
\mathbf{j}_{\varepsilon}^{\text{I}}.$ The heat current is connected
with the temperature gradient by $\mathbf{j}_{\varepsilon}^{\text{I}}%
=-\kappa\nabla T,$ with a positive-definite magnon thermal conductivity
$\kappa.$
%Parameters $\kappa$ and $R$ are calculated using a reasonable model
%later.
We then obtain%
\begin{equation}
\frac{d\left(  \Delta F_{\text{ne}}\right)  }{dt}=\int\kappa\frac{R\left(
T\right)  }{T}\left(  \nabla T\right)  ^{2}dx>0, \label{1overT}%
\end{equation}
for finite reflections of heat currents. The temperature dependence of the
free energy $\Delta F\left(  T\right)  $ is crucial to drive the DW
propagation. It has been shown that the equilibrium part $\Delta F_{\text{e}}$
decreases with an increasing temperature $T$ \cite{Nowak1,Nowak2,XR}, which
leads to a conclusion that the DW must move to the hotter region to reduce the
free energy due to the entropic force. However, the present system under the
temperature gradient $\nabla T$, strictly speaking, is not in an equilibrium
state. Moreover, the non-equilibrium part $\Delta F_{\text{ne}}$ monotonically
increases with time [for $R>0$ in Eq. \eqref{1overT}], thus the arguments
based on either maximization of entropy or minimization of free energy are not
always valid \cite{Nowak1,Nowak2,XR} since transport processes matter a lot in
non-equilibrium thermodynamics. In the following, we study the magnon
transport, in particular its backscatterings by a domain wall, and predict a
thermodynamic magnon recoil effect in the presence of a temperature gradient,
competing with the entropic force.

The thermal properties of magnons crucially depend on their dispersion
relations and lifetime. To calculate the magnon thermal conductivity $\kappa$,
we consider the heat current carried by the magnon flow due to the temperature
gradient $\mathbf{\nabla}T$ in the absence of domain walls (as the upper panel
in Fig. 1), $\mathbf{j}_{\varepsilon}^{\text{I}}=L^{-1}\sum_{k}\delta
n_{k}\hbar\omega\mathbf{v}_{g}\left(  \mathbf{k}\right)  ,$ where $L$ is the
wire length, $\mathbf{k}$ is the magnon wave-vector, $\delta n_{k}=n_{k}%
-\bar{n}_{k}$ is the magnon number in excess of equilibrium value $\bar{n}%
_{k}$ $=1/\left[  e^{\hbar\omega\left(  k\right)  /\left(  k_{B}T\right)
}-1\right]  $ being the Bose-Einstein distribution with Boltzman constant
$k_{B},$ $\hbar\omega\left(  k\right)  $ is the magnon energy, and
$\mathbf{v}_{g}\left(  \mathbf{k}\right)  =\partial\omega/\partial\mathbf{k}$
is the magnon group velocity. Using the Boltzmann approach we can write a
first-order expression for the excess magnon number in the steady state and in
the relaxation time approximation, $\delta n_{k}=-\tau_{k}\left(  \partial
\bar{n}_{k}/\partial T\right)  \mathbf{v}_{g}\cdot\mathbf{\nabla}T,$ where
$\tau_{k}$ is the magnon relaxation time. One thus obtains the magnon thermal
conductivity
\begin{equation}
\kappa=\frac{1}{2\pi}\sum_{n=1}^{N}\int_{\omega_{n}^{\min}}^{\omega_{n}^{\max
}}\tau_{k}\hbar\omega\left(  \partial\bar{n}_{k}/\partial T\right)
v_{g}d\omega, \label{Conductivity}%
\end{equation}
by using the one dimensional magnon density of states (DOS) $\rho\left(
k\right)  =L/2\pi.$ Here $N$ is the number of energy bands and $\omega
_{n}^{\min(\max)}$ is the lowest (highest) frequency of each band $n.$

The presence of domain wall may lead to a strong spin-wave reflection, and
thus a reduction of magnon heat currents $\mathbf{j}_{\varepsilon}^{\text{II}%
}=\left(  1-R\right)  \mathbf{j}_{\varepsilon}^{\text{I}}$. The reported mean
free path of thermal magnons in insulating ferromagnets, e.g., YIG, usually is $\sim1-100$ $\mu$m \cite{Ulrike,Heremans}, which is
much larger than the domain wall width $\delta_{D}\sim10-100$ nm, the
scattering of spin waves by the wall can thus be treated as a ballistic
process, thereby conserving the total momentum. We then can derive the
reflection probability $R$ of magnon heat currents by the wall via the
Landauer-B\"{u}ttiker formula \cite{Yan3,Loss}%
\begin{equation}
R\left(  T\right)  =\frac{\sum_{n=1}^{N}\int_{\omega_{n}^{\min}}^{\omega
_{n}^{\max}}F\left(  \omega,T\right)  \left\vert r\left(  k\right)
\right\vert ^{2}d\omega}{\sum_{n=1}^{N}\int_{\omega_{n}^{\min}}^{\omega
_{n}^{\max}}F\left(  \omega,T\right)  d\omega}, \label{Reflection}%
\end{equation}
where $r\left(  k\right)  $ is the $k-$dependent reflection coefficient of
magnons by the wall and $F\left(  \omega,T\right)  =\hbar\omega\left(
\partial\bar{n}_{k}/\partial T\right)  $. Here we do not consider the
modification of the magnon DOS due to the wall \cite{Yan3}, which is relevant
to reflectionless magnons treated in equilibrium thermodynamic theories
\cite{XR,Vindigni} but causes only negligible effects to our results here. In
the momentum-conserving scattering process between spin waves and the domain
wall, the change rate of the linear momentum of a DW is $dp_{\text{DW}%
}/dt=2\dot{\phi}M_{s}/\gamma$ \cite{Yan2} which must be compensated by that
from magnons (with wave vector $k$) $dp_{\text{magnons}}/dt=$ $\left(  \delta
n_{k}\right)  v_{g}\left\vert r\left(  k\right)  \right\vert ^{2}\hbar k$.
Here $\phi$ is the tilted angle of the DW plane, $M_{s}$ is the saturation
magnetization, and $\gamma$ is the gyromagnetic ratio. Spin-wave reflections
thus lead to a precession of the domain wall plane with the angular velocity
\begin{equation}
\dot{\phi}_{k}=\frac{\gamma}{2M_{s}}\left(  \delta n_{k}\right)
v_{g}\left\vert r\left(  k\right)  \right\vert ^{2}\hbar k. \label{Precession}%
\end{equation}
The equivalent magnetic field responsible for the above precession velocity is
then $\mathbf{h}_{k}=\dot{\phi}_{k}/\gamma,$ giving rise to an effective field
along the wire axis after a summation of all states
\begin{align}
\mathbf{H}_{\text{ne}}  &  =L^{-1}\sum_{k}\mathbf{h}_{k}\nonumber\\
&  =-\frac{\mathbf{\nabla}T}{4\pi M_{s}}\sum_{n=1}^{N}\int_{\omega_{n}^{\min}%
}^{\omega_{n}^{\max}}\tau_{k}\left(  \partial\bar{n}_{k}/\partial T\right)
v_{g}\left\vert r\left(  k\right)  \right\vert ^{2}\hbar kd\omega,
\label{Effectivefields}%
\end{align}
which is the effective field or force acting on the wall due to the
thermodynamic magnon recoil in a temperature gradient.

Equation \eqref{Effectivefields} is quite a general formula that can be used
to calculate the effective field under any magnon dispersion relations and
relaxation mechanisms. Experiment data in YIG, for instance, show an acoustic
branch with frequency that rises from nearly zero at the Brillouin zone center
to a value at the zone boundary that varies from $6$ to $9.5$ THz. These
values correspond to temperatures of approximately 300 and 500 K. Since the
lowest optical branch lies above the zone-boundary value, the calculation of
the thermal properties up to room temperature can be done considering only the
acoustic branch, i.e., $N=1$. At low wave numbers the dispersion relation can
be approximately by a quadratic form $\omega=\omega_{1}^{\min}+Jk^{2},$ where
$\omega_{1}^{\min}$ is the acoustic band gap depending on materials
parameters, such as the magnetic anisotropy, dipole-dipole coupling,
Dzyaloshiskii-Moriya interaction, etc., and $J$ is the exchange constant. The
cut-off frequency $\omega_{1}^{\max}$ thus is $\omega_{1}^{\max}=\omega
_{1}^{\min}+Jk_{\text{m}}^{2}$ with $k_{\text{m}}$ the maximum wave vector
depending on the magnon propagation direction. The magnon group velocity is
then $v_{g}=2\sqrt{J\left(  \omega-\omega_{1}^{\min}\right)  }$. It has been
shown that the quadratic dispersion agrees very well with the actual
dispersion up to a wave vector $k=0.6k_{\text{m}}$ in YIG \cite{Rezende}.
Under the above conditions, we obtain%
\begin{equation}
\mathbf{H}_{\text{ne}}=-\frac{\kappa\mathbf{\nabla}T}{2\pi M_{s}}%
\frac{\left\vert \bar{r}\right\vert ^{2}}{\bar{v}_{g}},
\label{Effectivefields2}%
\end{equation}
with the average reflection probability
\begin{equation}
\left\vert \bar{r}\right\vert ^{2}=\int_{\omega_{1}^{\min}}^{\omega_{1}^{\max
}}\tau_{k}F\left(  \omega,T\right)  \left\vert r\left(  k\right)  \right\vert
^{2}d\omega/\int_{\omega_{1}^{\min}}^{\omega_{1}^{\max}}\tau_{k}F\left(
\omega,T\right)  d\omega, \label{Averagereflection}%
\end{equation}
and the average group velocity
\begin{equation}
\bar{v}_{g}=\int_{\omega_{1}^{\min}}^{\omega_{1}^{\max}}\tau_{k}v_{g}F\left(
\omega,T\right)  d\omega/\int_{\omega_{1}^{\min}}^{\omega_{1}^{\max}}\tau
_{k}F\left(  \omega,T\right)  d\omega. \label{Averagegroupvelocity}%
\end{equation}
We then obtain the DW velocity along the direction of heat currents due to
their recoil effect, below Walker breakdown \cite{Walker},%
\begin{equation}
\mathbf{v}_{\text{ne}}\left(  T\right)  =\frac{\gamma\delta_{D}}{\alpha
}\mathbf{H}_{\text{ne}}=-\frac{\left\vert \bar{r}\right\vert ^{2}}{\bar{v}%
_{g}}\frac{\gamma\delta_{D}}{\alpha}\frac{\kappa\mathbf{\nabla}T}{2\pi M_{s}}.
\label{Velocity}%
\end{equation}

Schlickeiser \emph{et al.} \cite{Nowak2} derived an effective magnetic field
based on equilibrium thermodynamics, $\mathbf{H}_{\text{e}}=\mathbf{\nabla
}T\left(  J_{0}/T_{c}\right)  /\left(  \delta_{D}m_{\text{eq}}M_{s}a\right)  $
with nearest-neighbor exchange energy $J_{0}$, Curie temperature $T_{c},$
equilibrium local magnetization $m_{\text{eq}},$ and lattice constant $a,$
which is believed to exceed the magnonic spin transfer torque proportional to
$1-\left\vert \bar{r}\right\vert ^{2}$ \cite{Nowak2,Yan1}. So, the final DW
propagation direction depends on the competition between $\mathbf{H}%
_{\text{ne}}$ and $\mathbf{H}_{\text{e}}.$ The condition to observe a DW
propagation toward the colder regime is therefore%
\begin{equation}
\kappa\frac{\left\vert \bar{r}\right\vert ^{2}}{\bar{v}_{g}}>\frac{2\pi
}{\delta_{D}m_{\text{eq}}a}\left(  J_{0}/T_{c}\right)  ,\label{Condition}%
\end{equation}
which requires a good heat conduction in magnetic domains (a large $\kappa$
without the domain wall), a strong magnon backscattering (a large $\left\vert
\bar{r}\right\vert ^{2}$), a slow magnon group velocity (a small $\bar{v}_{g}%
$), and a broad domain wall (a large $\delta_{D}$). The forces induced by
non-equilibrium thermal fluctuations under temperature gradients cause a Brownian
motion \cite{Najafi,Kong} of the domain wall and could be another reason to
push its propagation toward the colder region \cite{SK}. However it is still
an open question how valid the classical fluctuation-dissipation theorem for
equilibrium states is \cite{Ma}, particularly when it is applied to nonequilibrium
steady states in the presence of temperature gradients
\cite{Chetrite,Solano,Prost,Speck}. The Brownian motion effect is however
negligible in the presence of strong magnon backscatterings.

In order to evaluate the parameters in criterion \eqref{Condition}, we now can
make either of two plausible assumptions about the behavior of $\tau_{k}.$
Model I: If one considers that the relaxation time $\tau_{k}$ is independent
of both the wave number and the temperature (the simple average-lifetime
model) \cite{Berger} and takes $\tau_{k}=\bar{\tau}_{k}$, we obtain
$\left\vert \bar{r}\right\vert ^{2}=R$ the same as the reflection probability
of magnon heat currents [Eq. \eqref{Reflection} when $N=1$], $\bar{v}_{g}%
=\int_{\omega_{1}^{\min}}^{\omega_{1}^{\max}}v_{g}F\left(  \omega,T\right)
d\omega/\int_{\omega_{1}^{\min}}^{\omega_{1}^{\max}}F\left(  \omega,T\right)
d\omega$, and $\kappa\propto\bar{\tau}_{k}$. Model II: If we consider the
Gilbert damping but neglect higher order processes such as magnon-magnon
interactions, the relaxation time is then $\tau_{k}=1/\left(  2\alpha
\omega\right)  $ with Gilbert damping constant $\alpha$ \cite{Gilbert}. We
thus have $\left\vert \bar{r}\right\vert ^{2}=\int_{\omega_{1}^{\min}}%
^{\omega_{1}^{\max}}\left(  \partial\bar{n}_{k}/\partial T\right)  \left\vert
r\left(  k\right)  \right\vert ^{2}d\omega/\int_{\omega_{1}^{\min}}%
^{\omega_{1}^{\max}}\left(  \partial\bar{n}_{k}/\partial T\right)  d\omega,$
$\bar{v}_{g}=\int_{\omega_{1}^{\min}}^{\omega_{1}^{\max}}\left(  \partial
\bar{n}_{k}/\partial T\right)  v_{g}d\omega/\int_{\omega_{1}^{\min}}%
^{\omega_{1}^{\max}}\left(  \partial\bar{n}_{k}/\partial T\right)  d\omega,$
and $\kappa\propto1/\alpha.$

It has been shown that both dipole-dipole \cite{Yan2} and Dzyaloshiskii-Moriya
\cite{Weiwei} interactions can result in strong magnon reflections in the
presence of a domain wall in ferromagnets. A precessing domain wall in
antiferromagnet can also lead to significant magnon reflections
\cite{Se,Bogdan}. A common feature of the reflection probability function
$\left\vert r\left(  k\left(  \omega\right)  \right)  \right\vert ^{2}$ is the
sharp transition from $1$ at lower frequencies to $0$ at higher frequencies
\cite{Yan2, Wang, Bogdan}, satisfying ansatz $\left\vert r\left(  k\left(
\omega\right)  \right)  \right\vert ^{2}=\left\vert w\left(  -\left(
\omega-\omega_{c}\right)  /\Delta\omega\right)  \right\vert ^{2},$ with the
transition frequency $\omega_{c}$ and the spectrum width $\Delta\omega.$
Function $\left\vert w\right\vert ^{2}$ reduces to $1$ for $\omega\ll
\omega_{c}$, and $0$ for $\omega\gg\omega_{c}.$ The form of the function $w$
depends on material parameters such as the domain wall width \cite{Yan3}, the
Dzyaloshiskii-Moriya interaction strength \cite{Weiwei}, etc., and scattering
details such as the incident angle of magnons \cite{Yan2}. However, for a very
narrow spectrum $(\Delta\omega\ll1)$ which is often the case
\cite{Yan2,Bogdan}$,$ it can be approximately described by the Heaviside step
function, i.e., $\left\vert w\right\vert ^{2}\approx s\left(  -\left(
\omega-\omega_{c}\right)  \right)  $ with
\begin{equation}
s\left(  x\right)  =\left\{
\begin{array}
[c]{c}%
0,\text{ }x<0\\
\frac{1}{2},\text{ }x=0\\
1,\text{ }x>0
\end{array}
\right.  . \label{Heaviside}%
\end{equation}
It indicates that magnons are completely reflected by the wall when their
frequencies are lower than $\omega_{c},$ while they essentially pass through
the domain wall without any reflection above $\omega_{c}.$ By denoting
$A=\hbar\omega_{1}^{\min}/k_{B}$, $B=\hbar\omega_{1}^{\max}/k_{B},$
$x_{a}=A/T,$ and $x_{c}=C/T$ with $C=\hbar\omega_{c}/k_{B},$ Eqs.
\eqref{Conductivity}, \eqref{Averagereflection}, and
\eqref{Averagegroupvelocity} can be calculated analytically and yield%

\begin{align}
\kappa &  =\frac{\bar{\tau}_{k}}{\pi}\sqrt{\frac{Jk_{B}^{5}T^{3}}{\hbar^{3}}%
}\int_{A/T}^{B/T}\frac{\sqrt{x-x_{a}}x^{2}e^{x}}{\left(  e^{x}-1\right)  ^{2}%
}dx,\label{M1k}\\
\bar{v}_{g}  &  =2\sqrt{\frac{Jk_{B}T}{\hbar}}\frac{\int_{A/T}^{B/T}%
\frac{\sqrt{x-x_{a}}x^{2}e^{x}}{\left(  e^{x}-1\right)  ^{2}}dx}{f\left(
A/T\right)  -f\left(  B/T\right)  },\label{M1vg}\\
\left\vert \bar{r}\right\vert ^{2}  &  =R=\frac{\int_{A/T}^{B/T}s\left(
-\left(  x-x_{c}\right)  \right)  \frac{x^{2}e^{x}}{\left(  e^{x}-1\right)
^{2}}dx}{f\left(  A/T\right)  -f\left(  B/T\right)  },
\label{ReflectionHeaviside}%
\end{align}
with \cite{Handbook}%
\begin{equation}
f\left(  x\right)  =-2x\ln\left(  1-e^{-x}\right)  +\frac{x^{2}}{e^{x}%
-1}+2\sum_{p=1}^{\infty}\frac{e^{-px}}{p^{2}}, \label{f}%
\end{equation}
in Model I $\left(  \tau_{k}=\bar{\tau}_{k}\right)  $, and%
\begin{align}
\kappa &  =\frac{1}{2\pi\alpha}\sqrt{\frac{Jk_{B}^{3}T}{\hbar}}\int%
_{A/T}^{B/T}\frac{\sqrt{x-x_{a}}xe^{x}}{\left(  e^{x}-1\right)  ^{2}%
}dx,\label{M2k}\\
\bar{v}_{g}  &  =2\sqrt{\frac{Jk_{B}T}{\hbar}}\frac{\int_{A/T}^{B/T}%
\frac{\sqrt{x-x_{a}}xe^{x}}{\left(  e^{x}-1\right)  ^{2}}dx}{g\left(
A/T\right)  -g\left(  B/T\right)  },\label{M2vg}\\
\left\vert \bar{r}\right\vert ^{2}  &  =\frac{\int_{A/T}^{B/T}s\left(
-\left(  x-x_{c}\right)  \right)  \frac{xe^{x}}{\left(  e^{x}-1\right)  ^{2}%
}dx}{g\left(  A/T\right)  -g\left(  B/T\right)  },
\label{ReflectionHeaviside2}%
\end{align}
with \cite{Handbook}
\begin{equation}
g\left(  x\right)  =-\ln\left(  1-e^{-x}\right)  +\frac{x}{e^{x}-1}, \label{g}%
\end{equation}
in Model II $\left(  \tau_{k}=1/\left(  2\alpha\omega\right)  \right)  .$
\begin{figure}[ptbh]
\begin{center}
\includegraphics[width=8.5cm]{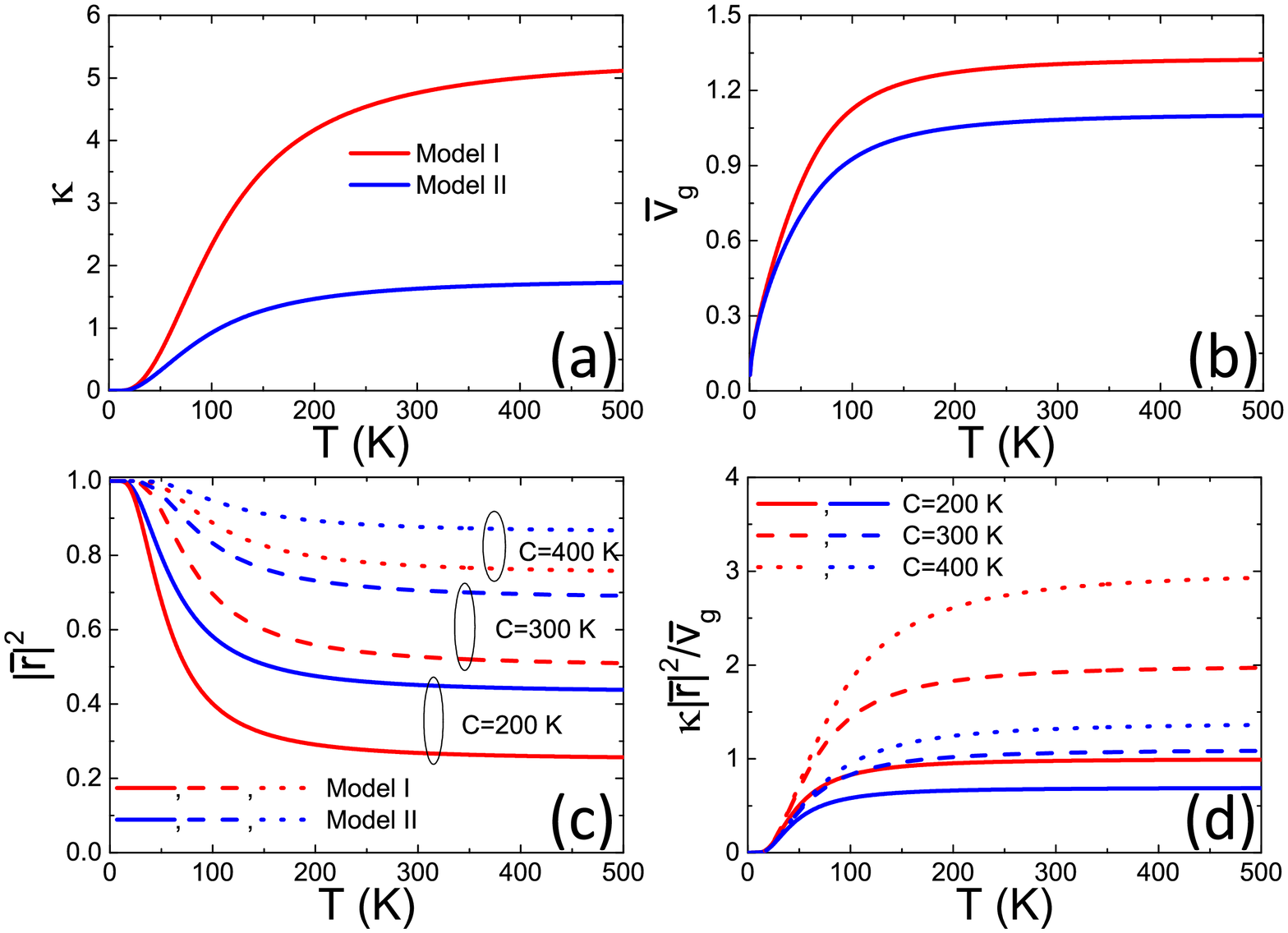}
\end{center}
\caption{(Color online) (a) Temperature dependence of the magnon heat
conductivity $\kappa$ normalized by $\bar{\tau}_{k}\sqrt{Jk_{B}^{5}A^{3}%
/\pi^{3}\hbar^{3}}$ in Model I, and by $\left(  2\pi\alpha\right)  ^{-1}%
\sqrt{Jk_{B}^{3}A/\hbar}$ in Model II, respectively. (b) Average group
velocity $\bar{v}_{g}$ as a function of temperature $T,$ in unit of
$2\sqrt{Jk_{B}A/\hbar}$ in two models. (c) Temperature dependence of average
reflection probability $\left\vert \bar{r}\right\vert ^{2}$ for different
transition temperatures $C$. (d) Parameter $\kappa\left\vert \bar
{r}\right\vert ^{2}/\bar{v}_{g}$ as a function of temperature for different
$C$, with unit of $\bar{\tau}_{k}k_{B}^{2}A/\left(  2\pi\hbar\right)  $ in
Model I and of $k_{B}/\left(  4\pi\alpha\right)  $ in Model II, respectively.
In the calculations, we use $A=100$ K and $B=500$ K. All temperatures are
below $T_{c}.$}%
\end{figure}

The following three cases are of potential interest. (i) For a low transition
frequency $\left(  \omega_{c}<\omega_{1}^{\min}\right)  ,$ there is no
reflection, and Eqs. \eqref{ReflectionHeaviside} and
\eqref{ReflectionHeaviside2} yield $\left\vert \bar{r}\right\vert ^{2}=0.$
(ii) For an intermediate transition frequency $\left(  \omega_{1}^{\min
}<\omega_{c}<\omega_{1}^{\max}\right)  ,$ Eqs. \eqref{ReflectionHeaviside} and
\eqref{ReflectionHeaviside2} reduce to $\left\vert \bar{r}\right\vert
^{2}=\left[  f\left(  A/T\right)  -f\left(  C/T\right)  \right]  /\left[
f\left(  A/T\right)  -f\left(  B/T\right)  \right]  $ and $\left[  g\left(
A/T\right)  -g\left(  C/T\right)  \right]  /\left[  g\left(  A/T\right)
-g\left(  B/T\right)  \right]  $, respectively. (iii) For a high transition
frequency $\left(  \omega_{c}>\omega_{1}^{\max}\right)  ,$ all magnons are
reflected by the wall. Thus, Eq. \eqref{ReflectionHeaviside} and
\eqref{ReflectionHeaviside2} reduce to $\left\vert \bar{r}\right\vert ^{2}=1.$

Figure 2(a) shows the temperature dependence of the magnon thermal
conductivity $\kappa$ in both Models I and II. It increases as the elevated
temperature in two cases. In Fig. 2(b) we calculate the average group velocity
$\bar{v}_{g}$. It monotonically increases with the temperature and saturates
at high temperatures. The temperature dependence of $\left\vert \bar
{r}\right\vert ^{2}$ for different transition temperatures $C$ are shown in
Fig. 2(c), with a monotonically decreasing manner. It is because  higher
temperatures make more magnons populate  higher energy levels, which leads
to a smaller magnon reflection subsequently. We also observe that a higher
transition temperature leads to a stronger magnon reflection. Figure 2(d)
demonstrates a monotonically increasing dependence on the temperature of
parameter $\kappa\left\vert \bar{r}\right\vert ^{2}/\bar{v}_{g}$ in both
models. As shown in Eq. \eqref{Condition} with quadratic magnon dispersion
relations, parameter $\kappa\left\vert \bar{r}\right\vert ^{2}/\bar{v}_{g},$
which is independent of exchange constant $J$, is crucial to determine if the
domain wall can move toward the colder region. According to our calculations,
this condition should be easily satisfied at elevated temperatures
($T\gtrsim400$ K) with a small magnon damping ($\bar{\tau}_{k}\gtrsim1$ ns or
$\alpha\lesssim10^{-4}$) for a broad domain wall ($\delta_{D}\gtrsim100$ nm)
in a weak ferromagnet ($J_{0}/a\lesssim3\times10^{-12}$ J/m) under any
temperature gradient that can overcome the pinning force produced by defects
or impurities. Other relaxation models considering three- and four-magnon
scattering processes are expected not to modify our conclusions significantly.

To summarize, we predict a thermodynamic magnon recoil effect for domain wall
motion under temperature gradients. We correct the previous
thermodynamic theories by including a heat current term for entropy and/or
free-energy generations, which is always presents in non-equilibrium steady
states in the presence of a temperature gradient. The heat current gets
modulations by the DW with momentum-conserving backscatterings. It then leads
to a recoil force on the wall, which competes with the previously identified
entropic force.
%The condition to observe a domain wall propagation toward the
%colder region is analytically obtained.
Our theory thereby closes the
inconsistency between macroscopic and microscopic theories for the domain wall
motion, and and we propose experiments to test it. We also expect the similar
thermodynamic magnon recoil effect to play an important role in other magnetic
structures, e.g., magnetic vortices, bubbles, or Skyrmions, and other
materials like antiferromagnets or multiferroics.

After the completion of this work, we became aware of one recent report \cite{SeKwon}
on a DW thermophoresis in antiferromagnets using the classical
fluctuation-dissipation relation without considering any magnon backscattering. Our results should also
be applicable to their work.

We thank Se Kwon Kim and Gerrit Bauer for useful discussions. This work is supported by DFG
Priority Programme 1538 \textquotedblleft Spin-Caloric
Transport\textquotedblright\ (P.Y. and J.S.)\ and European Union Seventh
Framework Programme \textquotedblleft SpinIcur\textquotedblright\ under Grant
No. FP7-People-2012-ITN-316657 (Y.C.).

\end{document}